\documentclass[onecolumn,amsmath,amsfonts,amssymb,showpacs]{revtex4} 
\DeclareRobustCommand{\orderof}{\ensuremath{\mathcal{O}}}
\usepackage[skip=-10pt]{caption}
\usepackage{graphicx}
\usepackage{float}
\usepackage{etoolbox}
\usepackage{mathptmx}      
\def\beq{\begin{equation}}
\def\eeq{\end{equation}}
\def\bea{\begin{eqnarray}}

\def\eea{\end{eqnarray}}

\begin{document}

\title{Exponential nonlinear electrodynamics and  backreaction effects on Holographic superconductor in the Lifshitz black hole background}
\author{Zeinab Sherkatghanad}
\email{z.sherkat@ph.iut.ac.ir}
\affiliation{Department of Physics, Isfahan University of Technology, Isfahan 84156-83111, Iran.}

\author{Behrouz Mirza}
\email{b.mirza@cc.iut.ac.ir}
\affiliation{Department of Physics, Isfahan University of Technology, Isfahan 84156-83111, Iran.}

\author{Fatemeh Lalehgani Dezaki}
\email{f2lalega@uwaterloo.ca}
\affiliation{Department of Physics and Astronomy, University of Waterloo,
Waterloo, Ontario, Canada, N2L 3G1}
\affiliation{Department of Physics, Isfahan University of Technology, Isfahan 84156-83111, Iran}

\begin{abstract}
We analytically describe the properties of the s-wave holographic superconductor with the Exponential nonlinear electrodynamics in the Lifshitz black hole background in four-dimensions. Employing an assumption the scalar and gauge fields backreact on the background geometry, we calculate the critical temperature as well as the condensation operator. Based on Sturm-Liouville method, we show that the critical temperature decreases with increasing exponential nonlinear electrodynamics and Lifshitz dynamical exponent, z, indicating that condensation becomes difficult. Also we find that the effects of backreaction has a more important role on the critical temperature and condensation operator in small values of Lifshitz dynamical exponent, while z is around one.
In addition, the properties of the upper critical magnetic field in Lifshitz black hole background using Sturm-Liouville approach is investigated to describe the phase diagram of the corresponding holographic superconductor in the prob limit. We observe that the critical magnetic field decreases with increasing Lifshitz dynamical exponent, z, and it goes to zero at critical temperature, independent of the Lifshitz dynamical exponent, z.
\end{abstract}
\maketitle


\section{Introduction}
The Anti-de Sitter/Conformal Field Theory (AdS/CFT) correspondence between a conformal field theory (CFT) in $d$ spacetime dimensions living on its boundary, and a theory of gravity in $(d+1)$-dimensional anti-de Sitter (AdS) spacetime is a novel idea where prepare a powerful tool for describing strongly coupled systems \cite{Maldacena,Aharony,Witten}.
The relation of theories of gravity and strongly interacting gauge theories in one lower dimensions, has arisen intensive investigations in order to study many different strongly interacting condensed matter systems \cite{Hartnoll,Herzog1,McGreevy,Herzog}.
Bardeen, Cooper and Schrieffer (BCS) propose the microscopic theory of superconductivity  that describes various properties of low temperature superconducting materials \cite{Bardeen,Bardeen1957}. But the mechanism of the high temperature superconductors has long been a mysteries problem in modern condensed matter physics.
Therefore, it was recently suggested that it is completely possible to describe the properties of the high temperature superconductors in condensed matter physics, by using a classical general relativity in one higher dimensional spacetime. The first atempt to investigate the properties of strongly coupled superconductors on the boundary field theory, named holographic superconductors, by a classical general relativity living in one higher dimension, was proposed by Hartnoll, et. al., \cite{Hartnoll1,Hartnoll2}.  Considering AdS/CFT correspondence to high temperature superconductors, admit a dual gravitational involving the mechanism of spontaneous breaking of Abelian gauge symmetry near the event horizon of the black hole.
Therefore, a charged scalar field and a Maxwell field  in the theory of gravity describe the scalar operator and the $U(1)$ symmetry in the dual field theory, respectively \cite{Gubser1,Gubser2}. In this condition, this holographic model experiences a phase transition from black hole with no hair (normal phase/conductor phase) to the scalar hair at low temperatures (superconducting phase).\\
\indent Recently, there are many efforts to describe different aspects  of the holographic superconductors
from different perspective \cite{Pan,Cai,Herzog,Franco,Mirza,Mahapatra,Mahapatra1,Fan,Zangeneh}. Also the effects of the nonlinear extension of the Maxwell electrodynamics as well as linear Maxwell field for holographic superconductors with backreaction effects have attracted a lot of attention \cite{Sheykhi,Jing,Wang,Hendi,Zangeneh1}. These calculations are based on both numerical  and analytical methods  e.g. the Sturm-Liouville (SL) eigenvalue problem and Matching method.

The presence of the AdS black holes in the gravity side in the holographic superconductor model can lead to a non-zero temperature for the boundary relativistic CFT.  The real condensed matter
systems are far from a relativistic one, so it is important to generalize these holographic superconducting models to non-relativistic cases \cite{Polchinski,Sin,Zhang,Bu}. \\
In \cite{Nishida} an extention of the AdS/CFT correspondence to non-relativistic conformal field (NR-CFT) was represented. Since the NR-CFT is invariant under Galilean transformations with Schrodinger symmetry,  they described holographic duals to Galilean conformal field theories with Schrodinger symmetry. Among other novel results, similar theories with Liftshitz symmetry is introduced and they find that non-relativistic CFT that describe multicritical points in certain magnetic materials and liquid crystals
can be dual to certain non-relativistic gravitational theories in the Lifshitz space-time
background \cite{Kachru}.

A non-relativistic renormalizable candidate of quantum gravity named by "Horava-Lifshitz (HL) theory" represented by Horava in \cite{Horava,Horava1}. The HL theory recast to Einstein’s general relativity at large distances and this theory assume that the Lorentz symmetry is broken in the ultraviolet. Additionally, the quantum phase transitions in many condensed matter systems are controled in terms of Lifshitz fixed points which render the anisotropic scaling of spacetime. In this way the anisotropic scalings between space and time as Lifshitz fixed points are given by \cite{Griffin}
\bea
t \to \lambda^z t \   ,  \    \  x \to \lambda x
\eea 
where z is the dynamical critical exponent

In the context of AdS/CFT correspondence one can show that many relativistic QFTs have relativistic
gravity duals, so it is natural to expect that the two disparate applications of Lifshitz scaling
nonrelativistic QFT on one hand and HL gravity on the other hand should similarly be related
by a holographic duality. Also Lifshitz spacetimes are vacuum solutions of HL gravity which suggests that HL gravity provides the minimal holographic dual for Lifshitz-type field theories with anisotropic scaling and dynamical exponent $z$ \cite{Griffin}. 
In Ref. \cite{Janiszewski,Janiszewski1} authors show a relation that connects non-relativistic (NR) QFTs, that is, many body quantum mechanical systems, to HL gravity. They investigated a duality between 
non-relativistic gravity theory (HL gravity) and any NR-CFTs which have the same
set of symmetry transformations such as time dependent spatial diffeomorphisms, spatially
dependent temporal diffeomorphisms, and the U(1) symmetry acting on the background
gauge field coupled to particle number. Thus, the natural arena for non-relativistic holography is non-relativistic HL gravity.\\

In the current paper, we are interested to consider the effects both exponential nonlinear electrodynamics with backreaction on Holographic superconductor in the Lifshitz black hole background. 
This paper is outlined as follows. Thus, analytically the properties of the s-wave holographic superconductor with the exponential nonlinear electrodynamics in the Lifshitz black hole background in four-dimensions is considered when the back reaction of the scalar and gauge fields are significant on the background geometry. The critical temperature as well as the condensation operator are investigated based on Sturm-Liouville method. Also we obtain that the effects of backreaction besides  exponential nonlinear electrodynamics, $\beta$ and Lifshitz dynamical exponent, $z$,  play an important role on the critical temperature and condensation operator. An interesting result represent that the effects of backreaction reveals in small values of Lifshitz dynamical exponent, z, while is  around one.
In addition, the properties of the upper critical magnetic field in the Lifshitz black hole background to describe the phase diagrame of the corresponding holographic superconductor is investigated. We observe that the critical magnetic field increases with increasing the Lifshitz dynamical exponent, z, for temperature smaller than the critical temperature and it goes to zero at critical temperature, independent of the Lifshitz dynamical exponent, z.\\

The paper is organized as follows:
In Section II, we investigate the field equations when the gauge field is in the form Exponential nonlinear electrodynamics and backreaction on $(3 + 1)-$holographic superconductor in the framework of planar AdS black holes in the Lifshitz black hole background is taken into acount. In Section III, in the presence of backreaction, the effects of exponential nonlinear electrodynamics parameter, $\beta$, and Lifshitz dynemical exponent, $z$ on critical temperature in terms of the charge density and on the condensations of the scalar operators is considered.  In Section IV,  the properties of the upper critical magnetic field in the Lifshitz black hole background to describe the phase diagrame of the corresponding holographic superconductor is investigated by using Sturm-Liouville method.

\section{EQUATIONS OF HOLOGRAPHIC SUPERCONDUCTORS WITH
BACKREACTIONS}
The action of Einstein gravity coupled to a charged complex scalar field in the presence of
exponential nonlinear electrodynamics with a negative cosmological constant in the Lifshitz background reads as follows \cite{Taylor}
\bea\label{action}
I=\int d^4 x \sqrt{-g}[\frac{1}{2 \kappa^2}(R-2 \Lambda-\frac{1}{2} \partial_{\mu} \Phi \partial ^{\mu} \Phi-\frac{1}{4} e^{\lambda \Phi} F_{\mu\nu}  F^{\mu\nu})+L_m],
\eea
here
\bea
&&L_m=L({\cal{F}})-\mid\bigtriangledown \psi- i \ q \ A\psi \mid^2-m^2 \mid\psi\mid^2\\\nonumber
&&L({\cal{F}})=\frac{1}{4 \beta^2} (e^{-\beta^2 {\cal{F}}}-1),
\eea
and $\Lambda$ is the cosmological constant
\bea
\Lambda=-\frac{(z+1)(z+2)}{2 l^2},
\eea
where $\kappa^2=8\pi G_4$ is the gravitational constant, $\beta$ represents the nonlinear parameter, $A$ is the gauge field, $\psi$ shows a scalar field with charge $q$ and mass $m$. Also, ${\cal{F}}= {\cal{F}}_{\mu\nu}   {\cal{F}}^{\mu\nu}$ in which ${\cal{F}}_{\mu\nu}$ is the electromagnetic field tensor. We adopt the following ansatz for the gauge and scalar fields, the Liftshitz plane-symmetric black hole with an asymptotically AdS behavior and taking the backreaction effects has the following form,
\bea\label{metric}
ds^2=-r^{2z} f(r) e^{-\chi(r)} dt^2+\frac{dr^2}{r^2 f(r)}+r^2(dx^2+dy^2).
\eea
The gauge choices for the vector field and the scalar field is considered in this way
\bea
A_{\mu}=(\phi(r),0,0,0) \  \  \   ,  \   \   \psi=\psi(r).
\eea
The Einstein equations of motion by varying the action, Eq. (\ref{action}) with respect to the metric is given by
\bea
\label{Einstein}
&&R^{\mu \nu}-\frac{g^{\mu \nu}}{2} R-\frac{(z+1)(z+2)}{2 l^2}g^{\mu \nu}-\frac{1}{2} \partial_{\mu} \Phi \partial _{\nu} \Phi-\frac{1}{2} e^{-\lambda \Phi} F_{\mu \rho} F_{\nu}^{\rho}\\\nonumber
&&+\frac{1}{8} g_{\mu\nu} e^{\lambda \Phi} F_{\mu\nu}  F^{\mu\nu}= \kappa^2 T^{\mu \nu},
\eea
here $T^{\mu \nu}$ is the energy momentum tensor,
\bea
T^{\mu \nu}&=&\frac{1}{4 \beta^2} g^{\mu \nu} (e^{-\beta^2 {\cal{F}}}-1)+e^{-\beta^2 {\cal{F}}} {\cal{F}}_{\sigma}^{\mu} {\cal{F}}^{\sigma \nu}-m^2 g^{\mu \nu} \mid\psi\mid^2\\\nonumber
&-&g^{\mu \nu} \mid\bigtriangledown \psi- iqA\psi \mid^2+[(\bigtriangledown^{\nu}- iqA^{\nu})\psi ^{\star} (\bigtriangledown^{\mu}- iqA^{\mu})\psi +\mu \leftrightarrow \nu].
\eea
The other equations of motion can be rewritten as follows
\bea
\label{FSC}
&&\bigtriangledown_{\mu}({\cal{F}}^{\mu \nu} e^{-\beta^2 {\cal{F}}})=iq(\psi^\star (\bigtriangledown^{\nu}-iq A^{\nu})\psi-\psi(\bigtriangledown^{\nu}+iq A^\nu)\psi^\star)\\
\label{PHSC}
&&(\bigtriangledown_{\mu}- iqA_{\mu})(\bigtriangledown^{\mu}- iqA^{\mu})\psi-m^2  \psi=0,
\eea
and
\bea
\label{FS}
&&\partial_{\mu} (\sqrt{-g} e^{\lambda \Phi} F^{\mu \nu})=0\\
&&\partial _{\mu}  (\sqrt{-g} \partial ^{\mu} \Phi)-\frac{\lambda}{4}  \sqrt{-g} e^{\lambda \Phi} F_{\mu \nu} F^{\mu \nu}=0.
\eea
In the probe limit, when we neglect the back-reaction effects in the equations of motion go back to $4$-dimensional Lifshitz black hole solution. In this case the only non-vanishing component of the field strength is $F_{rt}$ and we can obtain the following solution for Eq. (\ref{FS}),
\bea
\label{F}
F_{rt}=\hat{q} e^{-\lambda \Phi} r^{z-3}.
\eea
Also in this limit $\kappa \to 0$, combination of $tt$ and $rr$ components of the Einstein equations of motion, Eq. (\ref{Einstein}), give the following form,
\bea
\label{philif}
\partial_r \Phi \partial_r \Phi=\frac{4 (z-1)}{r^2},
\eea
where $\hat{q}$ is a charge which can be related to the scalar field $\Phi$
\bea
&&\hat{q}=2 l^2 (z-1)(z+2),\\
&&\lambda^2=\frac{4}{z-1}.
\eea
Therefor, by replacing Eq. (\ref{F}) and Eq. (\ref{philif}) into the the Einstein equations of motion, the Liftshitz plane-symmetric black hole in the probe limit turns into the following form
\bea\label{metricLSH}
ds^2=-r^{2z} (1-\frac{r_+^{z+2}}{r^{z+2}})  dt^2+\frac{dr^2}{r^2 (1-\frac{r_+^{z+2}}{r^{z+2}})}+r^2(dx^2+dy^2).
\eea
This is just the Lifshitz spacetime with non-trivial dilaton and gauge fields. When $\kappa = 0$ and $z=1$, the metric coefficient $f(r)$ recast into the case of Schwarzschild AdS black holes $f(r)= (1-\frac{r_+^3}{r^3})$.  Since in the presence of back-reaction, the scalar field $\phi$ and $\psi$ back react only on the background geometry, the relation of field strength $F_{rt}$ in Eq. (\ref{F}) and massless scalar field $\Phi$ in Eq. (\ref{philif}) do not change. If we replace the metric in Eq. (\ref{metric}) into the equations of motion Eqs. (\ref{Einstein}), (\ref{FSC}) and (\ref{PHSC}) we have the following set of equations,
\bea\label{chi}
\chi '(r)+2 \kappa ^2 r \left(\frac{\psi (r)^2 \phi (r)^2 e^{\chi (r)}}{f(r)^2 r^{2 z+2}}+\psi '(r)^2\right)=0,
\eea
\bea
\label{fn}
&&2 r f'(r)-r f(r) \  \chi '(r)+(2 z+4) f(r)-z^2 (1-e^{\chi (r)})+2 (1+e^{\chi (r)})+3 \ z-z \  e^{\chi (r)}\\\nonumber
&&+2 \kappa^2\Big(m^2 \psi (r)^2-\frac{1}{4 \beta ^2 }(e^{2 \beta ^2 e^{\chi (r)} r^{2-2 z} \phi '(r)^2}-1)+r^{2-2 z} \phi '(r)^2 \ e^{ 2 \beta ^2 e^{\chi (r)} r^{2-2 z} \phi '(r)^2+\chi (r)}\Big)=0,
\eea
\bea
\label{psi}
&&r^2 f(r) \psi ''(r)+\psi '(r) \left(r^2 f'(r)-\frac{1}{2}r^2 f(r) \chi '(r)+r (z+3) f(r)\right)\\\nonumber
&&+\psi (r) \left(\frac{e^{\chi (r)} \phi (r)^2}{f(r) r^{2 z}}-m^2\right)=0,
\eea
\bea
\label{phia}
&&r^2 f(r) \phi ''(r) \left(4 \beta ^2 e^{\chi (r)} r^{2-2 z} \phi '(r)^2+1\right)+2 \beta ^2 f(r) e^{\chi (r)} r^{4-2z} \chi '(r) \phi '(r)^3\\\nonumber
&&+2 \beta ^2 (2-2 z) f(r) e^{\chi (r)} r^{3-2 z} \phi '(r)^3+r^2 f(r) \phi '(r) \left(\frac{3-z}{r}+\frac{\chi '(r)}{2}\right)\\\nonumber
&&-2 \psi (r)^2 \phi (r) e^{-2 \beta ^2 e^{\chi (r)} r^{2-2 z} \phi '(r)^2}=0,
\eea
where $q=1$. We should note that  the effects of backreaction do not change Eq. (\ref{philif}) for scalar field $\Phi$  and also the non-vanishing component of the field strength, $F_{rt}$ can be calculated from Eq. (\ref{F}).  The above equations reduce to the equations in Ref. \cite{Wang} when exponential nonlinear parameter, $\beta \to 0$ and Lifshitz dynemical exponent, $z \to 1$ and they restore the equations in Ref. \cite{Sheykhi}, while $z \to 1$. The solutions of Eq. (\ref{psi}) and Eq. (\ref{phia}) in the limit of $r \to \infty$ which is correspond to the asymptotic behavior of the fields near the boundary, are given by
\bea\label{phiatboun}
&&\phi=\mu-\frac{\rho}{r^{d-z}},  \  \  \    z\neq d-2,\\
&&\phi=\mu-\rho \ \log r,  \  \  \    z=d-2,
\eea
and
\bea
&&\psi=\frac{\psi_-}{r^{\Delta_-}}+\frac{\psi_+}{r^{\Delta_+}},
\eea
where $d$ is the diemensions of space-time,  $\mu$ and $\rho$ are interpreted as the chemical potential and charge density in the dual field theory, respectively. Here we have,
\bea
\Delta_{\pm}=\frac{1}{2} \left(d-2+z\pm\sqrt{(d-2+z)^2+4 m^2}\right),
\eea
$\Delta $ is the conformal dimension of the dual operator $O_{\pm}$ in the boundary field theory. Also  $\psi_+$ and $\psi_-$
are taken into acount as the source and the vacuum expectation values of the dual operator. In the following we set $\psi_+$ equal to zero and investigate the condensation of $\psi_-=< O_- >$. \\
Hawking temperature of the black hole in the presence of backreaction effects at event horizon $r_+$ in which $f(r+) = 0$, is defined by
\bea\label{tempH}
T=\frac{r_+^{1+z} f'(r_+) e^{\chi(r_+)/2}}{4 \pi},
\eea
here $f'(r_+)$ and $\chi(r_+)$ can be calculated from Eqs. (\ref{chi}) and (\ref{fn}), by investigating the asymptotic behavior of the fields at the event horizon,
\bea
&&\phi(r_+)=0,\\
&&\psi(r_+)=\frac{f'(r_+) \psi'(r_+)}{m^2},
\eea
thus we get,
\bea
\label{frp}
f'(r_+)&=&\frac{z^2}{2 r_+} (1-e^{\chi (r_+)})-\frac{2}{2 r_+}  (1+e^{\chi (r_+)})-\frac{3 \ z}{2 r_+} +\frac{z \  e^{\chi (r_+)}}{2 r_+}\\\nonumber
&-&\frac{\kappa^2}{r_+}  \ \Big(m^2 \psi (r_+)^2-\frac{1}{4 \ \beta ^2 }(e^{2 \beta ^2 e^{\chi (r_+)} r_+^{2-2 z} \phi '(r_+)^2}-1)\\\nonumber
&+&r_+^{2-2 z} \phi '(r_+)^2 \ e^{ 2 \beta ^2 e^{\chi (r_+)} r_+^{2-2 z} \phi '(r_+)^2+\chi (r_+)}\Big).
\eea
Throughout this paper, we adopt the AdS radious $l=1$ and 
the scalar mass $m^2 = -2$ and $d=4$, and hence the corresponding dual operator has the mass dimension equals $\Delta_{\pm}=\frac{1}{2} \left(2+z\pm\sqrt{(2+z)^2-8}\right)$. 

In the next section we first review the Sturm-Liouville method and investigate the analytical properties of a (2+1)-holographic superconductor in the presence of exponential nonlinear electrodynamics. Then we obtain the critical temperature as function of the charge density and scalar condensation by the effects of both backreaction and exponential nonlinear electrodynamics in Lifshitz background.

\section{ THE CRITICAL TEMPRETURE AND CONDENSATION OF THE SCALAR
OPERATOR}
Now let us consider the Sturm-Liouville method to analytically calculate the critical temperature with respect to the charge density and describe the scalar condensation near the critical temperature. Thus, based on the Sturm-Liouville method we may describe the effects of backreaction and exponential nonlinear electrodynamics in Lifshitz background. Using the variable $u = \frac{r_+}{r}$, the Einstein equations of motion and scalar field $\phi$, electrodynamic field equations turn into
\bea\label{chi1}
&&\chi '(u)-\frac{2 \kappa ^2 r_+^2}{u^3 f(u)^2} \left(\frac{u^4 f(u)^2 \psi '(u)^2}{r_+^2}+e^{\chi (u)} \psi (u)^2 \phi (u)^2 \left(\frac{r_+}{u}\right){}^{-2 z-2}\right)=0,
\eea
\bea
\label{f1}
&&-2 u^2 f'(u)-u^2 f(u) \chi '(u)+u (2 z+1) f(u)+3 u f(u)-u z^2 \left(1-e^{\chi(u)}\right)\\\nonumber
&&+2 u \left(e^{\chi(u)}+1\right)+3 z u-z ue^{\chi(u)}-\frac{\kappa ^2 u}{2 \beta ^2} (e^{\frac{2 \beta ^2 u^4 e^{\chi (r)} \left(\frac{r_+}{u}\right){}^{2-2 z} \phi '(u)^2}{r_+^2}}-1)\\\nonumber
&&+\frac{2 \kappa ^2 u^4 \left(\frac{r_+}{u}\right){}^{1-2 z}}{r_+} \phi '(u)^2e^{\frac{2 \beta ^2 u^4 e^{\chi (u)} \left(\frac{r_+}{u}\right){}^{2-2 z} \phi '(u)^2}{r_+^2}+\chi (u)}+2 \kappa ^2 m^2 u \psi (u)^2=0,
\eea
\bea
\label{psi1}
&&u^2 f(u) \psi ''(u)-u (1+z) f(u) \psi '(u)-m^2 \psi (u)+u^2 f'(u) \psi '(u)\\\nonumber
&&-\frac{1}{2} u^2 f(u) \chi '(u) \psi '(u)+\frac{e^{\chi (u)} \psi (u) \phi (u)^2 \left(\frac{r_+}{u}\right){}^{-2 z}}{f(u)}=0,\\\nonumber
\eea
\bea
\label{phi1}
&&u^2 f(u) \phi ''(u)+\frac{1}{2} u^2 f(u) \chi '(u) \phi '(u)+u (z-3) f(u) \phi '(u)-2 \psi (u)^2 \phi (u)  \\\nonumber
&&\times e^{-\frac{2 \beta ^2 u^4 e^{\chi (u)} \left(\frac{r_+}{u}\right){}^{2-2 z} \phi '(u)^2}{r_+^2}}+\frac{2 \beta ^2 u^6 f(u)}{r_+^2} e^{\chi (u)} \left(\frac{r_+}{u}\right){}^{2-2 z} \chi '(u) \phi '(u)^3\\\nonumber
&&+\frac{4 \beta ^2 u^6 f(u)}{r_+^2} e^{\chi (u)} \left(\frac{r_+}{u}\right){}^{2-2 z} \phi '(u)^2 \phi ''(u)+\frac{8 \beta ^2 u^5 f(u)}{r_+^2} e^{\chi (u)} \left(\frac{r_+}{u}\right){}^{2-2 z} \phi '(u)^3\\\nonumber
&&+\frac{4 \beta ^2 u^4 (z-1) f(u)}{r_+} e^{\chi (u)} \left(\frac{r_+}{u}\right){}^{1-2 z} \phi '(u)^3=0.
\eea
we perturbatively solve the above Eqs. with the Sturm-Liouville approach. In what follows we calculate the effects of backreaction and exponential nonlinear electrodynamics perturbatively. The scalar operator $\epsilon=<O_{\pm}>$ is introduced as an expansion parameter while its value is small, close to the critical point. In this condition we can expand the gauge field $\phi(u)$, the scalar field $\psi(u)$, the metric functions $f(u)$ and $\chi(u)$ in the following form,
\bea\label{expand1}
&&\psi(u)=\epsilon \psi_1(u)+\epsilon^3 \psi_3(u)+\epsilon^5 \psi_5(u)+......
\eea
\bea\label{expand2}
&&\phi(u)= \phi_0(u)+\epsilon^2 \phi_2(u)+\epsilon^4 \phi_4(u)+......
\eea
\bea\label{expand3}
&&f(u)= f_0(u)+\epsilon^2 f_2(u)+\epsilon^4 f_4(u)+......
\eea
\bea\label{expand4}
&&\chi(u)=\epsilon^2 \chi_2(u)+\epsilon^4 \chi_4(u)+......
\eea
The above expansion and $\chi'(u)$ in Eq. (\ref{chi1}) are replaced into the Einstein equations of motion for metric function and scalar field equation that are defined in Eqs. (\ref{f1}) and (\ref{phi1}) respectively. Now keep the leading terms upto the first order of $\epsilon$ and second order of $\beta$ we have
\bea
\label{f}
&&2 u (z+2) f_0(u)-2 u^2 f_0'(u)+3 \beta ^2 \kappa^2 u^5 \left(\frac{r_+}{u}\right)^{-4 z} \phi_0'(u)^4\\\nonumber
&&+\kappa^2 u^3 \left(\frac{r_+}{u}\right)^{-2 z} \phi_0'(u)^2-2 u (z+2)=0,
\eea
\bea
\label{phi}
&&u^2 \phi_0''(u)+u (z-1)  \phi_0'(u)+4 \beta ^2 u^3 \left(\frac{r_+}{u}\right)^{-2 z}  \Big(u \phi_0'(u)^2 \phi_0''(u)\\\nonumber
&&+(z+1) \phi_0'(u)^3\Big)=0.
\eea
These equations can be solved perturbatively up to the second order of $\beta$. The metric function and the scalar field are obtained as follows
\bea
&&f_0(u)= \Big(1 -u^{z+2}+\frac{1}{2 (z-2)} C_1 ^2 \kappa ^2 r_+^{-2 z} (u^{z+2}- u^{4})\Big)+\frac{\beta ^2}{ (z-2) z} \\\nonumber
&&\times \Big(\frac{ C_1^4 \kappa ^2   }{3 z+2}(z-2) (z-1) r_+^{4-8 z} (u^{z+2}- u^{4 z+4})+\frac{15 C_1^4 \kappa ^2 (z-2) z r_+^{-4 z} (u^{ z+2}-u^{8})}{10 (z-6)}\\\nonumber
&&+\frac{2 C_1^4 \kappa ^2 (u^{8 z}-u^{ z+2}) (z-3) z }{ 10 r_+^4 (z-2)}\Big)+O(\beta^3),
\eea 
and 
\bea\label{scalar1}
\phi_0(u)=\frac{C_1 \left(u^{2-z}-1\right)}{2-z}-\frac{2 \beta ^2 C_1^3}{(6-z) \  r_+^{2 z}} \  (u^{6-z}-1)+O(\beta^3)  \   \   ,    \   \  z \ne 2.
\eea
where $C_1$ is a constant and is determined by equating the derivatives of Eqs. (\ref{phiatboun}) and (\ref{scalar1}) at the boundray ($u=0$), 
\bea
 C_1=-\frac{(2-z) \rho}{r_{+c}^{2-z}}.
\eea
If we substitute $\chi'(u)$ from Eq. (\ref{chi1}) into the electrodynemic field equation $\psi(u)$ in Eq. (\ref{psi1}) and keep the expansion upto the second order of $\epsilon$ and $\beta$ we get
\bea\label{psib}
&&\psi_1''(u) u^2 f_0(u) + \psi_1'(u) u^2 f_0'(u)-\psi_1'(u)u (z+1) f_0(u) \\\nonumber
&&+\frac{\psi_1(u) \phi_0(u)^2 \left(\frac{r_+}{u}\right)^{-2 z}}{f_0(u)}-m^2 \psi_1(u)=0,
\eea
here the behavior of $\psi_1$ near the asymptotic AdS boundary is given by
\bea\label{psiNB}
\psi_1(u)=\frac{<O_i>}{\sqrt{2} r_+^{\Delta_i}} u^{\Delta_i} F(u),
\eea
where, $F(u)$ is a variational trial function near the boundary which satisfy in the boundary condition $F(0)=1$ and $F'(0)=0$. Now the electrodynamic field equation for $\psi_1(u)$ in (\ref{psib})
turns into the following form
\bea
&&F''(u) f_0(u) u^{2 \Delta -z-1}+F'(u)  u^{2 \Delta -z-2} \left(u f_0'(u)-f_0(u) (-2 \Delta +z+1)\right)\\\nonumber
&&+F(u) u^{2 \Delta -z-3} \Big(\Delta \  u f_0'(u)+\Delta \  f_0(u) (\Delta -z-2)\\\nonumber
&&-m^2+\frac{C_1^2  \left(u^2-u^z\right)^2 r_+^{-2 z}}{(z-2)^2 f_0(u)}+\frac{4 \beta ^2 C_1^4  \left(u^{2-z}-1\right) }{ (z-2) \ (6-z) \ f_0(u) \ r_+^{2 z}}(u^{6-z}-1) \left(\frac{r_+}{u}\right)^{-2 z}\Big)=0.
\eea
We can rewrite the above equation as
\bea
T F''+T' F'+P F+\Gamma^2 Q F=0,
\eea
where
\bea
&&T=f_0(u) u^{2 \Delta -z-1},
\eea
\bea
&&P=u^{2 \Delta -z-3} (\Delta \  u f_0'(u)+\Delta \  f_0(u) (\Delta -z-2)-m^2),
\eea
\bea
Q=u^{2 \Delta -z-3}\Big[\frac{4 \ (z-2) \ \beta ^2 \ C_1^2  \left(u^{2-z}-1\right) }{ (6-z) f_0(u)}(u^{6-z}-1)  \left(\frac{r_+}{u}\right)^{-2 z}+\frac{ \left(u^2-u^z\right)^2 }{f_0(u)}\Big],
\eea
and $\Gamma= -\frac{C_1}{r_{+c}^z (2-z)}$. According to the Sturm-Liouville eigenvalue problem \cite{Sheykhi}, the eigenvalue $\Gamma^2$ can be determined by minimizing the expression
\bea\label{gamma}
\Gamma^2=\frac{\int_0^1 (T  F'^2 - P F^2)}{\int_0^1 Q F^2},
\eea
as a guess we have considered the trial function in the form 
\bea
F(u)=1-\alpha u^2,
\eea
where, $\alpha$ is a constant. In addition the critical temperature is defined as
\bea\label{TC1}
T_c=\frac{r_{+c}^{1+z} f'(r_{+c})}{4 \pi}
\eea
keep the terms upto the first order of $\epsilon$ and second order of $\beta$, $f'(r_{+c})$ in Eq. (\ref{frp}) can be turn into
\bea\label{fp}
f'(r_{+c})=\frac{z+2}{r_{+c}}-\frac{3}{2} \beta ^2 \kappa^2 r_+^{3-4 z} \phi_0'(r_{+c})^4-\frac{1}{2}\kappa^2 r_{+c}^{1-2 z} \phi_0'(r_{+c})^2.
\eea
Throughtout our calculation we are interested to see the effects of backreaction approximation perturbatively and upto the second order. Thus we keep the backreaction parameter upto the second order $\kappa^2$ and ignore the higher order terms. Substitute Eq. (\ref{scalar1}) and Eq. (\ref{fp}) into the critical temperature, Eq. (\ref{TC1}) and keep the terms upto the second order of $\beta$ we get,
\bea
T_c=r_{+c}^z \Big(z+2+\frac{1}{2} \beta ^2 \kappa^2  \Gamma^4 (2-z)^4-\frac{1}{2} \kappa^2 \Gamma ^2 (2-z)^2\Big).
\eea
the critical tempreture, $T_c$ can be calculated from $\Gamma^2$ in Eq. (\ref{gamma}) by itteration method. In order to use the itteration procedure in our calculations, we express the backreaction parameter as
\bea
\kappa=\kappa_n=n \ \Delta \kappa,
\eea
where $n=0,1,2,..., n_{max}$ and $\Delta \kappa=\kappa _{n+1}-\kappa_n$ is the step size of iterative method. Therefore, using the iterative procedure we define the following expressions:
\bea
\kappa^2 \Gamma^2=\kappa_n^2 \Gamma^2=\kappa_n^2 \Gamma ^2_{\kappa_{n-1}}++\orderof\left(\Delta \kappa^4 \right).
\eea 
It should be noted that $\Gamma^2$ is defined upto the second order of $\beta$, thus we have
\bea
&&\beta^2 \Gamma^2=\beta^2 \Gamma^2_{\beta^2=0}+\orderof\left( \beta^4\right)\\
&&\beta^2 \kappa^2 \Gamma^4=\beta^2 \kappa^2 _n \Gamma^4 _{ \kappa_{n-1} ,\beta^2=0}+\orderof\left( \beta^4\right)+\orderof\left(\Delta \kappa^4 \right).
\eea 
We start our calculation with the first step when $n=0$ or the backreaction parameter $\kappa_n$ is equal to zero and consider $\kappa_{-1}=0$, $\Gamma_{\kappa_{-1}}=0$ and $\Gamma_{\kappa_{-1},\beta^2=0}=0$, it means that at the first step we stand at the probe limit. Also we have $\Gamma_{\beta^2=0}$ equal to the values of $\Gamma^2$  in Eq. (\ref{gamma}) when $\beta^2=0$. In the second step, $n=1$, the backreaction parameter has a small value equal to $\kappa_1=\Delta \kappa$ and we can replace $\Gamma_{\kappa_{0}}$ and $\Gamma_{\kappa_{0},\beta^2=0}$ from the first step. Totally, at each step $\Gamma_{\kappa_{n-1}}$ and $\Gamma_{\kappa_{n-1},\beta^2=0}$ can be calculated from previous one.
Inserting Eq. (\ref{fp}) into the critical temperature and consider the itteration procedure we have
\bea\label{temp}
Tc&=&\frac{\sqrt{\left(\frac{\rho }{\Gamma }\right)^z} }{4 \pi }\Big(z+2+\frac{1}{2} \beta ^2 \kappa^2 _n \Gamma^4 _{ \kappa_{n-1} ,\beta^2=0} (2-z)^4-\frac{1}{2} \kappa_n^2 \Gamma ^2_{\kappa_{n-1}} (2-z)^2\Big),
\eea
\begin{figure}
\centering
  \includegraphics[width=7.5cm,height=9.5cm]{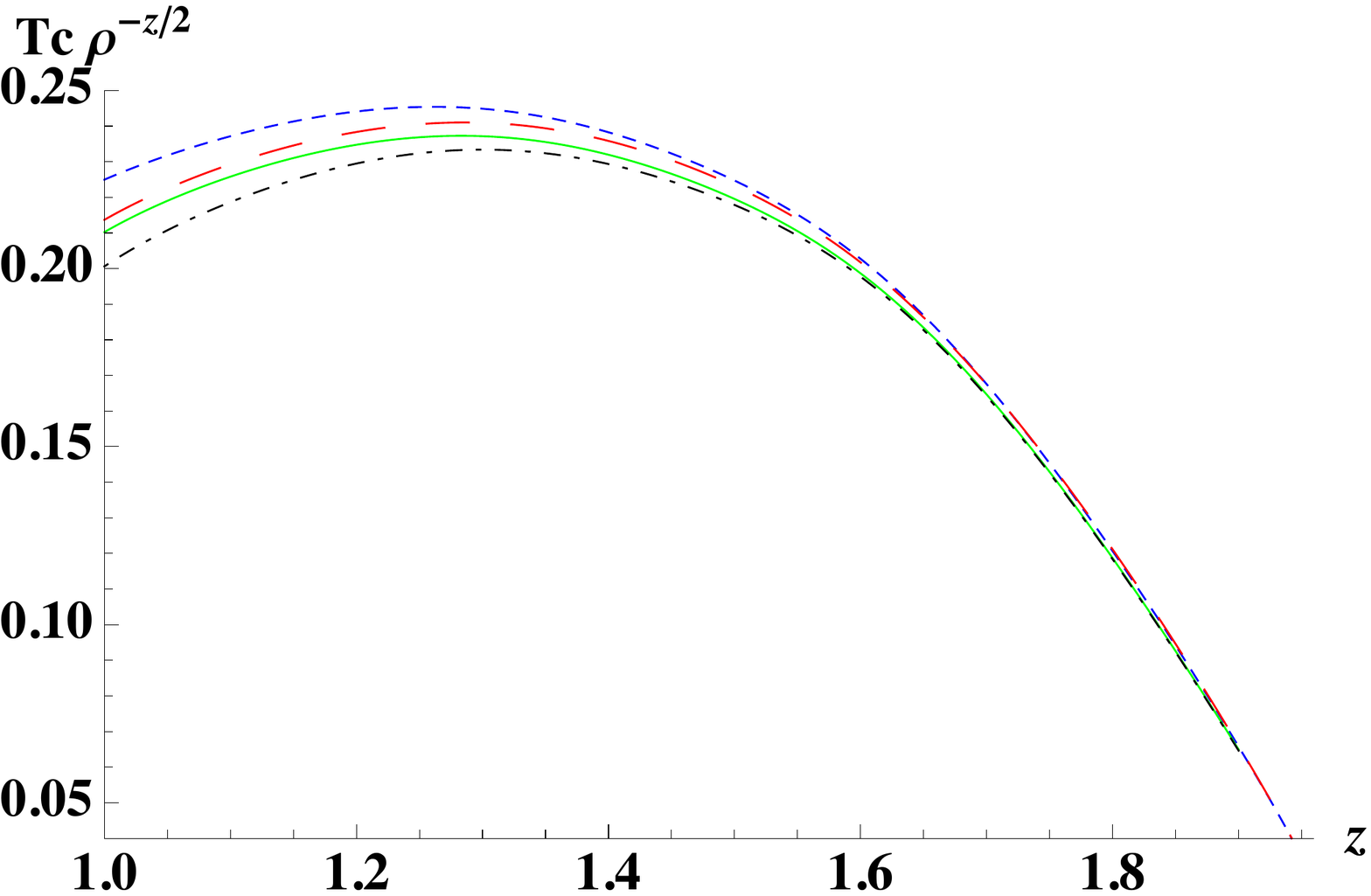}\\
  \caption {\it{$T_c \ \rho^{-z/2}$ with respect to the Lifshitz dynamical exponent $z$ for  $\kappa=0, \beta = 0$, $\kappa=0.56, \beta = 0$, $\kappa=0, \beta = 0.4$ and $\kappa=0.56, \beta = 0.4$ which are correspond to dotted blue, dashed red, solid green and dotted black lines, respectively. }}\label{figure:GTBHRN}
 \end{figure}

\begin{figure}
\centering
  \includegraphics[width=7.5cm,height=9.5cm]{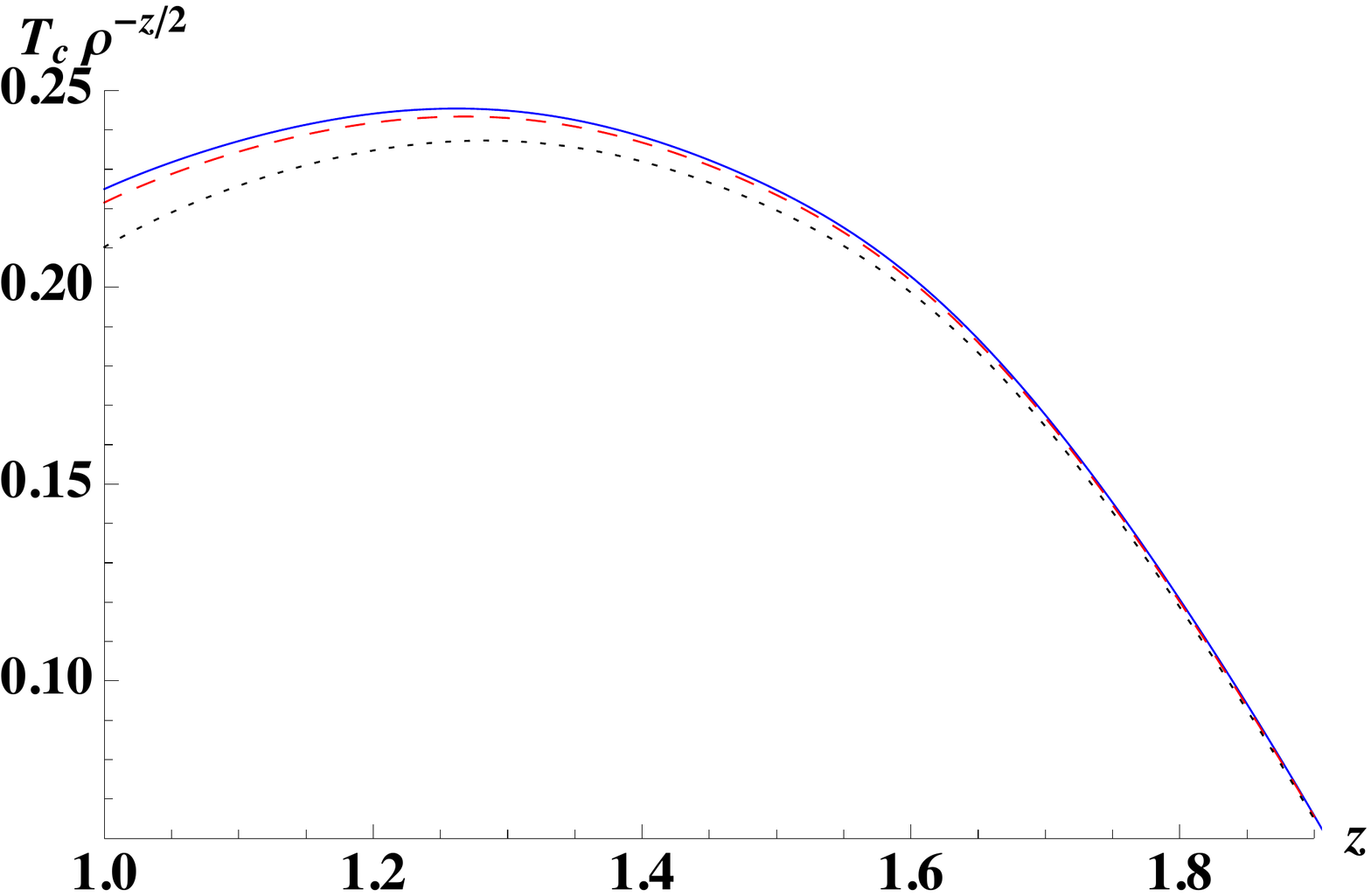}\\
  \caption {\it{The critical temperature over $\rho^{z/2}$ with respect to the Lifshitz dynamical exponent $z$ for  $\kappa=0$ and different values of $\beta = 0, 0.2, 0.4$ which are correspond to solid blue, dashed red and dotted black lines, respectively. }}\label{figure:GTBHRN}
 \end{figure}
\noindent here $\frac{\rho }{\Gamma }=r_{+c}^2$ 
and we neglect Higher order terms $\orderof\left( \beta^4\right)$ and $\orderof\left(\Delta \kappa^4 \right)$.
By introducing the itteration procedure, we are in a position to
use the analytical approach and obtain the critical temperature $T_c$ for a (2 + 1)-dimensional holographic superconductors in the presence of exponential nonlinear electrodynamics and backreaction in Lifshitz background.\\
Since the nonlinear parameter $\beta$ and backreation parameter $\kappa$ are taken small enough, we depict the critical temperature over $\rho^{z/2}$ with respect to the Lifshitz dynamic $z$ for  $\beta = 0, 0.1, 0.3$, $m^2=-2$ with the steps of $\Delta \kappa=0.04$ upto $n_{max}=14$  in Fig. 1. 
We plot $\frac{T_c}{\rho^{z/2}}$ with respect to $z$ for $\kappa=0$ and $\beta = 0, 0.2, 0.4$ in Fig. 2. The results show that the critical temperature decreases with increasing exponential nonlinear electrodynamics and Lifshitz dynamical exponent, z, indicating that condensation becomes difficult while backreaction is neglected, $\kappa=0$. Also, the effects of backreaction play an important role on the critical temperature for small values of Lifshitz dynamical exponent, z, while it is  around one. It means that the scalar fiels backreact on the gravity when we are not far from relativistic limit (Lifshitz dynamics parameter $z$ close to one). Meanwhile, for a fixed value of the nonlinear parameter $\beta$, the critical temperature drops as the backreaction parameter $\kappa$, increases for z around one. Thus, we conclude that the critical temperature becomes smaller and make the condensation harder when we increase the values of both backreaction and nonlinear parameters. In the following based on the Sturm-Liouville method as described above, we analytically derive the scalar condensation and the order of the phase transition with backreactions and exponential nonlinear electrodynamics near the critical temperature in the Lifshitz black hole background.\\
By using the definition of scalar field $\psi_1(u)$ near the boundary in Eq. (\ref{psiNB}), the equation of motion (\ref{phi1}) when temperature is close to the critical temperature can be expressed as,
\bea\label{CP}
&&u^{2-z} \phi ''(u)+(z-1) u^{1-z} \phi '(u)+\beta ^2 \Big(4 \ (z+1) r_+^{-2z} u^{3+z} \phi '(u)^3\\\nonumber
&&+4 \ u^{4+z} \phi '(u)^2 \phi ''(u) r_+^{-2z}\Big)+<O_->^2 r_+^{-2 \Delta } \Big(\frac{1}{2} \kappa ^2  F'(u)^2 u^{2 \Delta -z+3} \phi '(u)\\\nonumber
&&+\Delta \  \kappa ^2 F(u)  F'(u) u^{2 \Delta -z+2} \phi '(u)+\frac{\kappa ^2 F(u)^2 \phi (u)^2 \left(\frac{r_+}{u}\right){}^{-2 z} u^{2 \Delta -z} \phi '(u)}{2 f_0(u)^2}\\\nonumber
&&-\frac{F(u)^2  \phi (u) u^{2 \Delta -z}}{f_0(u)}+\frac{1}{2} \Delta ^2 \kappa ^2 F(u)^2  u^{2 \Delta -z+1} \phi '(u)\Big)+\orderof(\beta^4)=0,
\eea
since we are close to the critical temperature $T_c$, the parameter $<O_-> r_+^{-2 \Delta } $ is small enough. By using Eq. (\ref{scalar1}) we can expand $\phi(u)$ perturbatively in the following form
\bea\label{CP1}
\phi(u)=\frac{C_1 \left(u^{2-z}-1\right)}{2-z}-\frac{2 \beta ^2 C_1^3}{(6-z) \  r_+^{2 z}} \  (u^{6-z}-1)+<O_->^2 r_+^{-2 \Delta }  {\cal{G}}(u),
\eea
If replace the scalar field from Eq. (\ref{CP1}) into Eq. (\ref{CP}) we have
\bea
 r_+^{-2 \Delta } u^{3-2 z} \zeta'(u)+4 \beta ^2 C_1^2  u^{1- 2 z} r_+^{-2 \Delta -2 z}  \left(\zeta(u) \ u^{6}\right)'=\delta(u)+ \beta ^2 \eta(u)
\eea
where
\bea
&&\zeta=u^{z-1} {\cal{G}}'(u)\\
&&\delta=\frac{C_1 u^{-4 z} \left(\frac{r_+}{u}\right){}^{-2 z} r_+^{-2 (\Delta +z)}}{2 (z-6) (z-2)^2 f_0(u){}^2} \Big(C_1^2 \kappa ^2 (z-6) F(u)^2 r_+^{2 z} \left(u^{2 \Delta +1}\right) \left(u^2-u^z\right)^2 \\\nonumber
&&+\kappa ^2 (z-6) (z-2)^2 \ f_0(u){}^2 \ r_+^{2 z} \ F'(u)^2 \left(\frac{r_+}{u}\right){}^{2 z} u^{2 (\Delta +z)+4}\\\nonumber
&&+2 \Delta  \kappa ^2 (z-6) (z-2)^2\  f_0(u){}^2 \ F(u) \ r_+^{2 z} \ F'(u) \left(\frac{r_+}{u}\right){}^{2 z} u^{2 \Delta +2 z+3}\\\nonumber
&&+\Delta ^2 \kappa ^2 (z-6) (z-2)^2 f_0(u){}^2 F(u)^2 \ r_+^{2 z} \left(\frac{r_+}{u}\right){}^{2 z} u^{2 \Delta +2 z+2}\\\nonumber
&&-2 (z-6) (z-2) \ f_0(u) \ F(u)^2 \ r_+^{2 z} \left(u^z-u^2\right) \left(\frac{r_+}{u}\right){}^{2 z} u^{2 \Delta +2 z}\Big)
\eea
\bea
&&\eta=\frac{1}{2 (z-6) (z-2)^2 f_0(u){}^2} \Big( C_1 u^{-4 z} \left(\frac{r_+}{u}\right){}^{-2 z} r_+^{-2 (\Delta +z)} \Big(-2 \  C_1^2 \kappa ^2 (z-6)\\\nonumber
&&\times (z-2)^2 f_0(u){}^2 F'(u)^2 \left(\frac{r_+}{u}\right){}^{2 z} u^{2 (\Delta +z)+8}\\\nonumber
&&-4 C_1^2 \Delta  \kappa ^2 (z-6) (z-2)^2 f_0(u){}^2 F(u) F'(u) \left(\frac{r_+}{u}\right){}^{2 z} u^{2 \Delta +2 z+7}\\\nonumber
&&-2 C_1^2 \Delta ^2 \kappa ^2 (z-6) (z-2)^2 f_0(u){}^2 F(u)^2 \left(\frac{r_+}{u}\right){}^{2 z} u^{2 \Delta +2 z+6}\\\nonumber
&&+4 C_1^2 (z-2)^2 f_0(u) F(u)^2 \left(u^z-u^6\right) \left(\frac{r_+}{u}\right){}^{2 z} u^{2 \Delta +2 z}\\\nonumber
&&+16 C_1^2 u^6 (z-6) (z-2)^2 f_0(u){}^2 r_+^{2 \Delta } \left(r_+^{2 z}-u^{2 z} \left(\frac{r_+}{u}\right){}^{2 z}\right)\\\nonumber
&&+2 C_1^4 \kappa ^2 F(u)^2 u^{2 \Delta +1} \left(u^2-u^z\right) \left((z-6) u^{z+4}+2 (z-2) u^z+u^6 (10-3 z)\right)\Big)\Big)
\eea
\begin{figure*}
\centering
\begin{tabular}{cc}
\rotatebox{0}{
\includegraphics[width=0.47\textwidth,height=0.32\textheight]{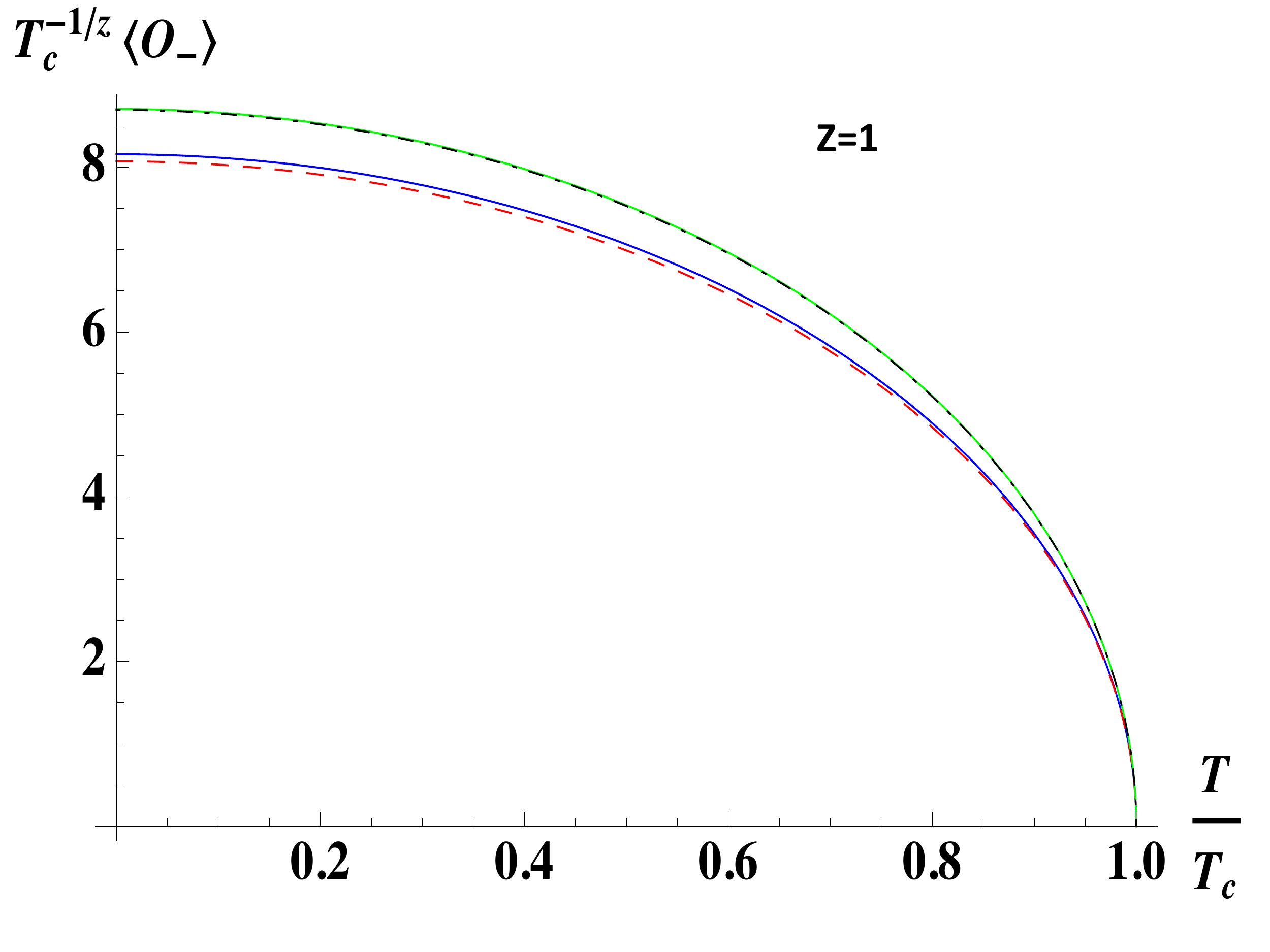}}&
\rotatebox{0}{
\includegraphics[width=0.45\textwidth,height=0.3\textheight]{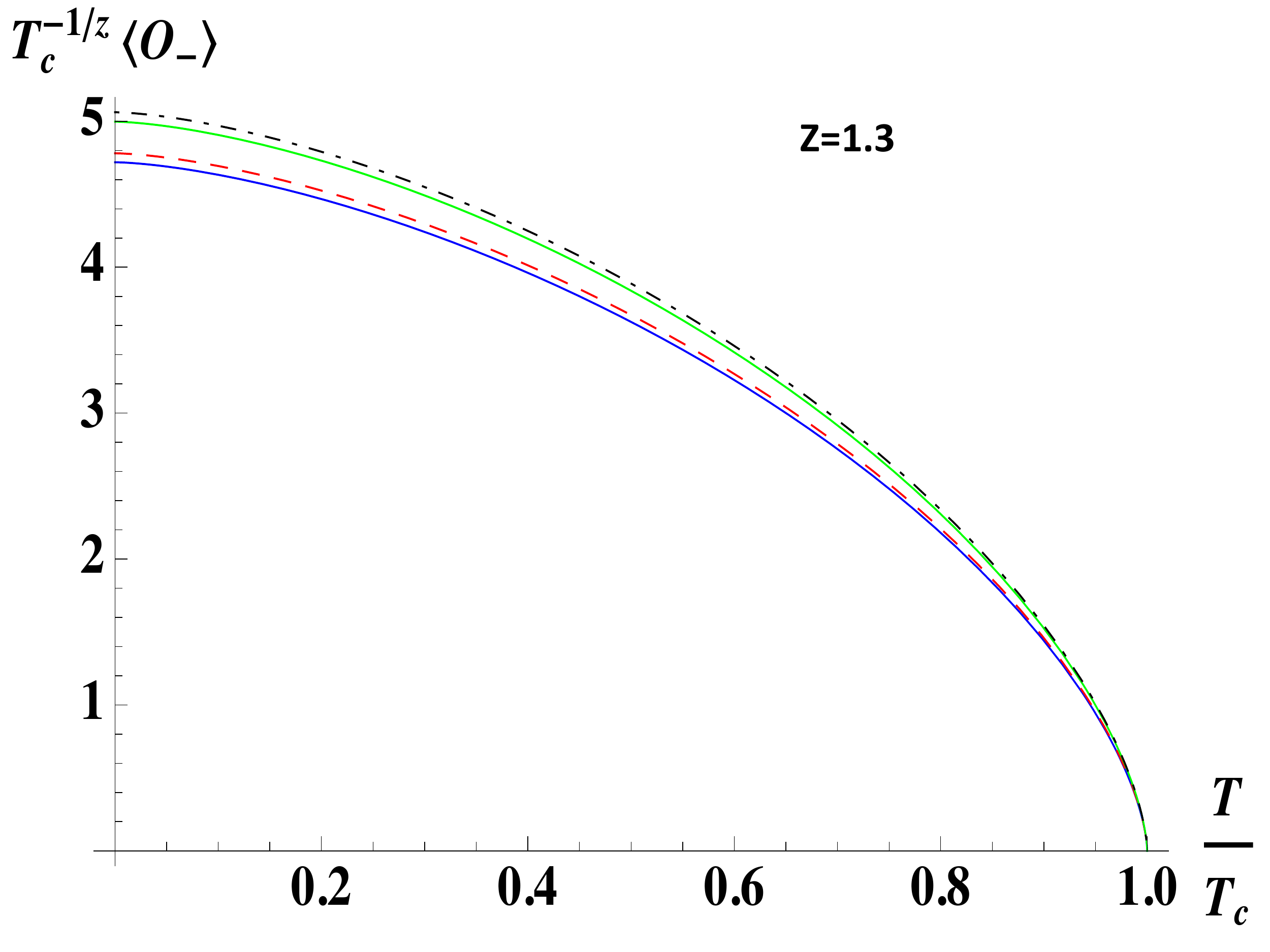}}\\
\rotatebox{0}{
\includegraphics[width=0.47\textwidth,height=0.32\textheight]{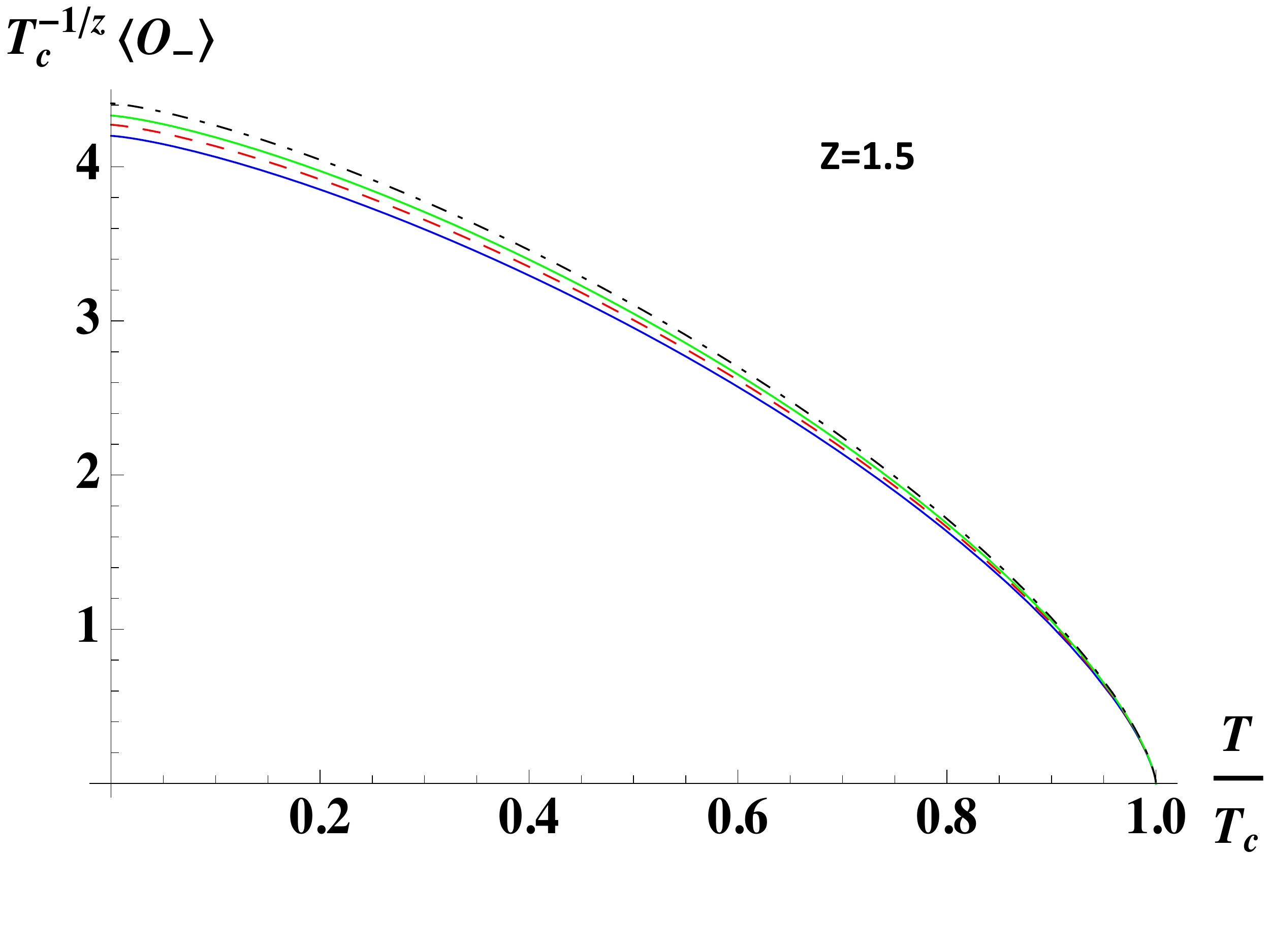}}&
\rotatebox{0}{
\includegraphics[width=0.47\textwidth,height=0.32\textheight]{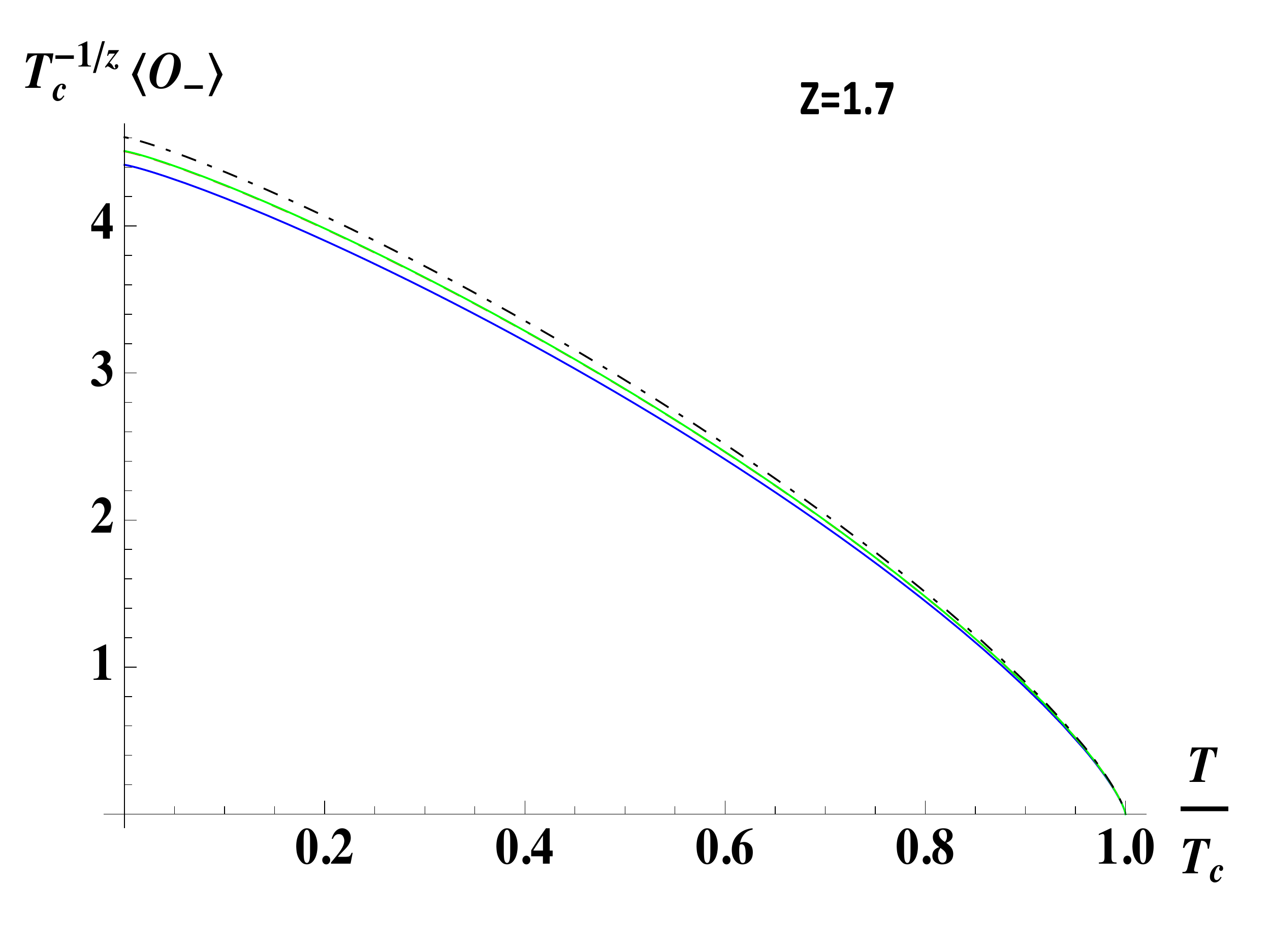}}\\
\\
\end{tabular}
\caption{ \it{The values of $<O_-> T_c^{-1/z}$ as a function of temperature in each diagram we have  $\kappa=0, \beta = 0$ (dashed red  line); $\kappa=0.56, \beta = 0$ (solid blue  line); $\kappa=0, \beta = 0.4$ (dot dashed black  line) and $\kappa=0.56, \beta = 0.4$ (solid green line). }
}\label{figure:Tbr}
\end{figure*}

In this condition, we can obtain an expression  for $u^{z-1} {\cal{G}}'(u)$ perturbetively upto the second order of $\beta$ by using the following boundray conditions,
\bea
{\cal{G}}'(1)={\cal{G}}(1)=0,
\eea
Therefore we get,
\bea
{\cal{G}}'(0)=-\frac{1}{u^{z-1}}\Big(\int_0^1 \frac{\delta(u)}{r_+^{-2 \Delta } u^{3-2 z}} \, du-\beta ^2 \int_0^1 \frac{u^{2 z-3} r_+^{2 \Delta -2 z} \left(2 C_1^2 u^5 \delta (u)-(z-1) r_+^{2 z} \eta (u)\right)}{z-1} \, du\Big)
\eea
The asymptotic behavior of the fields near the boundary for $z \ne d-2$ is given by
\bea
\phi(u)=\mu-\frac{\rho}{r_+^{2-z}} u^{2-z},
\eea
Equationg the above equation and Eq. (\ref{CP1}) we have 
\bea
\mu-\frac{\rho}{r_+^{2-z}} u^{2-z}&=&\frac{C_1 \left(u^{2-z}-1\right)}{2-z}\left(1-2\beta ^2 C_1^2 (\frac{(2-z)  \left(u^{6-z}-1\right)}{(6- z) r_+^{2 z+4} \left(u^{2-z}-1\right)})\right)\\\nonumber
&+&\frac{<O>^2}{ r_+^{2 \Delta }}  ({\cal{G}}(0)+u \  {\cal{G}}'(0)+...)
\eea
here the second line represents the expansion of ${\cal{G}}(u)$ near the boundray. If we consider the coefficients of the $u^{2-z}$ to be equal together we arrive
\bea
-\frac{\rho}{r_+^{2-z}}&=&\frac{C_1}{2-z}+\frac{<O_->^2}{ r_+^{2 \Delta }} u^{z-1} {\cal{G}}'(0)\\\nonumber
&=&\frac{C_1}{2-z}-\frac{<O_->^2}{ r_+^{2\Delta }}  \Big[\int_0^1 \frac{u^{2 z-3} r_+^{2 \Delta -2 z} \left(2 C_1^2 u^5 \delta (u)-(z-1) r_+^{2 z} \eta (u)\right)}{z-1} \, du\Big]
\eea
From Eqs. (\ref{tempH}) and (\ref{temp}) we can obtain
\bea
T_c^{2/z}-T^{2/z}&=&\frac{\frac{-\rho (2-z) \ r_+^z }{C_1}-r_+^2}{(4 \pi )^{2/z}}\\\nonumber
&\times&\Big[z+2+\frac{1}{2} \beta ^2 \kappa^2 _n \Gamma^4 _{ \kappa_{n-1} ,\beta=0} (2-z)^4-\frac{1}{2} \kappa_n^2 \Gamma ^2_{\kappa_{n-1}} (2-z)^2\\\nonumber
&+&\beta ^2\kappa^2 _n \Gamma^4 _{ \kappa_{n-1} ,\beta=0} (z-2)^3 r_+^{-4 z-4} \left(z r_+^{8 z}-3 r_+^{8 z}+r_+^8 z-3 r_+^8 +\frac{2 r_+^8}{z}\right)\Big]^{2/z}\\\nonumber
&=&\frac{<O_->^2(z-2) r_+^z}{(4 \pi )^{2/z} C_1}  \times \Big(z+2+\frac{1}{2} \beta ^2 \kappa^2 _n \Gamma^4 _{ \kappa_{n-1} ,\beta=0} (2-z)^4\\\nonumber
&-&\frac{1}{2} \kappa_n^2 \Gamma ^2_{\kappa_{n-1}} (2-z)^2 \Big)^{2/z}\\\nonumber
&\times&\Big[\int_0^1 \frac{\delta(u)}{u^{3-2 z}} \, du-\beta ^2 \int_0^1 \frac{u^{2 z-3} r_+^{2 \Delta -2 z} \left(2 C_1^2 u^5 \delta (u)-(z-1) r_+^{2 z} \eta (u)\right)}{z-1} \, du\Big]
\eea
Therefore the expectation value of $<O_->$ near the critical point is 
\bea
<O_->=\gamma \ T_c ^{1/z} \left(1-(\frac{T}{T_c})^{2/z}\right)^{\frac{1}{2}}
\eea
and $\gamma$ is the condensation parameter of the system,
\bea
\gamma&=&\frac{<O_-> \sqrt{2/z}}{\sqrt{T_c^{2/z}-T^{2/z}}}\\\nonumber
&=&\frac{(4 \pi )^{1/z} \sqrt{\frac{2 C_1}{z (z-2) r_+^z}} \Big(z+2+\frac{1}{2} \beta ^2 \kappa^2 _n \Gamma^4 _{ \kappa_{n-1} ,\beta=0} (2-z)^4-\frac{1}{2} \kappa_n^2 \Gamma ^2_{\kappa_{n-1}} (2-z)^2 \Big)^{-1/z}}{\sqrt{\Big[\int_0^1 \frac{\delta(u)}{u^{3-2 z}} \, du-\beta ^2 \int_0^1 \frac{u^{2 z-3} r_+^{2 \Delta -2 z} \left(2 C_1^2 u^5 \delta (u)-(z-1) r_+^{2 z} \eta (u)\right)}{z-1} \, du\Big]}}
\eea
Therefore, the results imply that the exponential nonlinear coupling, backreaction parameter and Lifshitz dynamical exponent do not change the the critical exponent. Thus phase transition for the superconductor belongs to the second order and the critical exponent of the system has the mean-field value 1/2.\\
The diagram of $<O_-> T_c^{-1/z}$ with respect to $T/T_c$ is depicted in Fig. 3. As we can see 
 the backreaction parameter $\kappa$ and $\beta$ play an important role on the condensation $<O_->$ while temperature is far from the critical temperature $T<T_c$. Also by increasing the Lifshitz dynamic $z$ for the fixed parameter $\kappa$, $<O_-> T_c^{-1/z}$ will be decreased, it mens that the condensation becomes difficult.
Meanwhile, the backreaction effects on $<O_-> T_c^{-1/z}$  by increasing the  Lifshitz dynamic $z$ will be neglected. Thus the backreaction effects are significant while $z$ is close to one.

\section{ External Magnetic Field in the Lifshitz black hole  background}
The external magnetic field in the holographic superconductors on the boundary corresponds to magnetic field in the bulk, according to the gauge/gravity duality. In this section we investigate the effects of Lifshitz dynamic on an external static magnetic field based on Sturm-Liouville method in the probe limit in which the scalar and gauge fields do not affect the background metric. We can make the following ansats
\bea
&&A_\mu dx^{\mu}=\phi (u) dt+(B y) dx\\\nonumber
&&\psi=\psi(u,y)
\eea
In this condition equation of motion for the scalar field $\psi$, Eq. (\ref{psib}) can be rewriten in the following form 
\bea\label{BF}
&&\psi _1''(u,y)-(1+z) \frac{\psi_1'(u,y)}{u}+\frac{\phi_0(u)^2 r_+^{-2 z} \psi_1(u,y)}{u^{2-2 z} f_0(u)^2}+\frac{2 \psi_1(u,y)}{u^2 f_0(u)}\\\nonumber
&&+\frac{f_0'(u) \psi_1'(u,y)}{f_0(u)}+\frac{\partial _y^2 \psi _1(u,y)-B^2 y^2 \psi_1(u,y)}{r_+^2 f_0(u)}=0
\eea
here prime is derivative with respect to $u$. To solve the above equation we consider $\psi(u,y)=R(u) Y(y)$. Replace this separable form for $\psi(u,y)$ into Eq. (\ref{BF}) we have 
\bea\label{BF1}
&&R''(u)-\frac{z \ R'(u)}{u}-\frac{R'(u)}{u}-\frac{B R(u)}{r_+^2 f_0(u)}+\frac{2 R(u)}{u^2 f_0(u)}+\frac{f_0'(u) R'(u)}{f_0(u)}\\\nonumber
&&+\frac{R(u) \phi _0(u){}^2 \left(\frac{r_+}{u}\right)^{-2 z}}{u^2 f_0(u){}^2}=0
\eea 
here we take the $y$ dependent part as the quantum harmonic oscillator in one dimension
with the relevant frequency determined by the magnetic field, $Y''(y)+B^2 y^2 Y(y)=c_n B \ Y(y)$, where, $c_n = 2n + 1$ is a constant. At the boundary, the asymptotic solution of Eq. (\ref{BF1}) is $R(u) = J_i \ u^{\Delta_i}$ where $i=+,-$. In this part we set $J_- = 0$ and study the asymptotic behaviour of $R(u)$ near the AdS boundray
\bea\label{BF2}
R(u)=J_+ u^{\Delta_+} \Theta(u) 
\eea
similar to previous disscution we can replace Eqs. (\ref{scalar1}) and (\ref{BF2}) into Eq.(\ref{BF1}) and find an equation in this form
 \bea
K \Theta''+K' \Theta'+P \Theta+\lambda^2 Q \  \Theta=0
\eea
where
\bea
&&K=f_0(u) u^{2 \Delta -z-1}
\eea
\bea
&&P=-\frac{B u^{2 \Delta -z-1}}{r_+^2}+\Delta  f_0'(u) u^{2 \Delta -z-2}+\Delta ^2 f_0(u) u^{2 \Delta -z-3}-\Delta  z f_0(u) u^{2 \Delta -z-3}\\\nonumber
&&-2 \Delta f_0(u) u^{2 \Delta -z-3}+2 u^{2 \Delta -z-3}
\eea
\bea
&&Q=\frac{ u^{2 \Delta +z-3}}{ f_0(u)}-\frac{2  u^{2 \Delta -1}}{ f_0(u)}+\frac{ u^{2 \Delta -z+1}}{f_0(u)}
\eea
where $\Gamma= -\frac{C_1}{r_{+c}^z (2-z)}$. According to the Sturm-Liouville eigenvalue problem, the eigenvalue $\lambda^2$ can be determined by minimizing the expression
\bea
\Gamma^2=\frac{\int_0^1 (K  \Theta'^2 - P \Theta^2)}{\int_0^1 Q \Theta^2}
\eea
we choose $\Theta$ in the following form 
\bea
\Theta (u)=1-\alpha \ u^2
\eea
where $\alpha$ is a constant. Minimizing $\Gamma$ we obtain the critical magnetic field and its dependence on Lifshitz dynamical exponent. Since the answers for the critical magnetic field are too long we plot $\frac{B_c}{T_c^{2z}}$ as a function of $T/T_c$ for different values of Lifshitz dynamic $z$ in Fig. 4. We find that, by increasing the nonlinear parameter, $1<z<d$ the critical magnetic field decreases.It means that the critical magnetic field is more stable for $z=1$. Here we considered the probe limit in which the scalar and gauge fields do not affect the background
metric.\\
These results are consistent with the phenomenology and the Ginzburg-Landau theory of superconductivity, it means that the critical magnetic field $B_c$  vanishes linearly close to critical temperature, $T \to T_c$ and it increases with $T_c$. 
\begin{figure}
\centering
  \includegraphics[width=7.5cm,height=10cm]{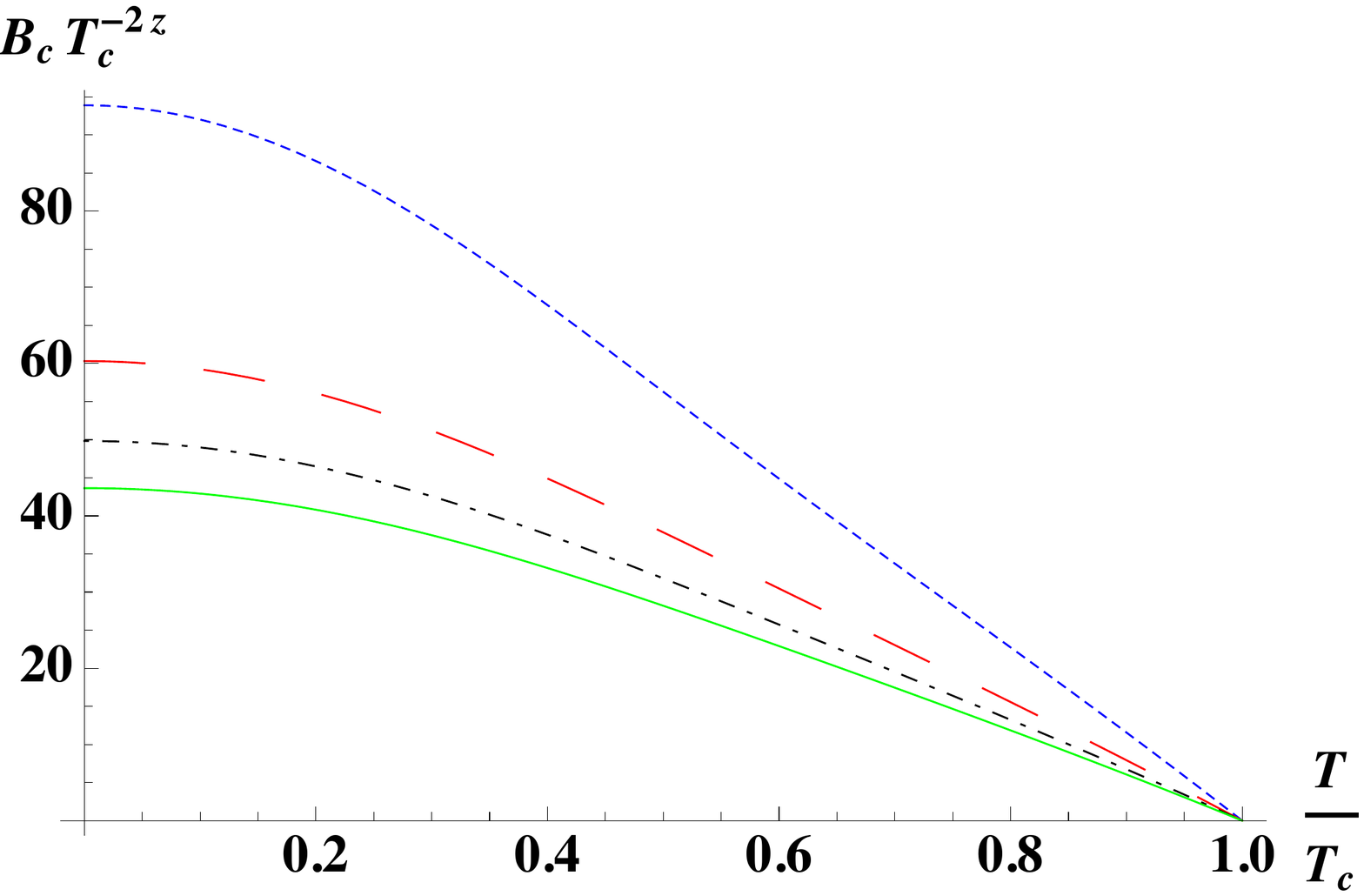}\\
  \caption {\it{$\frac{B_c}{T_c^{2z}}$ with respect to  $\frac{T}{T_c}$ in the prob limit,  $\kappa=0$, for different values of $z =1,1.3,1.5,1.7$ correspond to dotted blue, dashed red and dotdashed black and solid green lines, respectively }}\label{figure:GTBHRN}
 \end{figure}

\section{Conclusion}
In this work we explored the properties of holographic superconductors by putting the Abelian-Higgs model (s-wave) with Einstein-nonlinear electrodynamics as the gauge field into Lifshitz black hole geometry.  We assumed a limit in which the scalar and gauge fields backreact on the background metric. For this purpose we have employed the Sturm-Liouville analytic
method to investigate the problem. Therefore, the holographic superconductor model was constructed in the presence of the gravitational action with exponential nonlinear electrodynamics in Lifshitz black-hole space-times when the backreaction effect become stronger.\\
\indent It is found that the nonlinear corrections to the gauge filed as well as the backreaction effects and Lifshitz dynamical exponent play an important role on the critical temperature and the process of the scalar field condensation. In addition, the condensation of the scalar hair on the boundary becomes harder in the presence of both nonlinear electrodynamics and Lifshitz dynamical exponent. Increasing the Lifshitz dynamical exponent from $z=1$ upto the $z=3/2$ in four dimensional space-time increases $T_c \rho^{-z/2}$ and decreases $<O_-> T_c^{-1/z}$. Then for $3/2<z<2$, $T_c \rho^{-z/2}$ decreased and $<O_-> T_c^{-1/z}$ parameter increased. Meanwhile, we observed that the backreaction affects on the critical temperature and condensation operator in smaller values of Lifshitz dynamical exponent, while $z$ is around one. This implies that both the nonlinear corrections to the gauge field as well as backreaction and Lifshitz dynamical exponent, lead the formation of condensation harder. The results show that the critical exponent has the mean-field value $1/2$ by putting both exponential nonlinear electrodynamics and backreaction effects in the Lifshitz background in the holographic superconductor model. \\
At the end, the effect of Lifshitz dynamical exponent of the holographic superconductor on an external static magnetic field was investigated by adding a magnetic field in the bulk in the prob limit based on Sturm-Liouville analytic method. The results revealed the dependence of the critical magnetic field $B_c$ on parameters  $z$. In this case, the critical magnetic field values decrease with increasing $z$ for temperature smaller than the critical temperature, indicating that the critical magnetic field is more stable at $z=1$.

\section*{Acknowledgment}

This work is supported by Iranian National Science Foundation (INSF). We would like to thank department physics of Waterloo University for warm hospitality.


\end{document}